\documentstyle[epsfig,12pt,a4p]{article}

\newcommand{\pipipi}{\mbox{$\pi^0\pi^0\pi^0$ }}
\newcommand{\pipipic}{\mbox{$\pi^+\pi^-\pi^0$ }}

\begin{document}
\begin{titlepage}
\def\footnoterule{\hrule width 1.0\columnwidth}
\begin{tabbing}
put this on the right hand corner using tabbing so it looks
 and neat and in \= \kill
\> {10 April 2001}
\end{tabbing}
\bigskip
\bigskip
\begin{center}{\Large  {\bf A study of the
centrally produced \pipipi channel
in pp interactions at 450 GeV/c}
}\end{center}
\bigskip
\bigskip
\begin{center}{        The WA102 Collaboration
}\end{center}\bigskip
\begin{center}{
D.\thinspace Barberis$^{  4}$,
F.G.\thinspace Binon$^{   6}$,
F.E.\thinspace Close$^{  3,4}$,
K.M.\thinspace Danielsen$^{ 11}$,
S.V.\thinspace Donskov$^{  5}$,
B.C.\thinspace Earl$^{  3}$,
D.\thinspace Evans$^{  3}$,
B.R.\thinspace French$^{  4}$,
T.\thinspace Hino$^{ 12}$,
S.\thinspace Inaba$^{   8}$,
A.\thinspace Jacholkowski$^{   4}$,
T.\thinspace Jacobsen$^{  11}$,
G.V.\thinspace Khaustov$^{  5}$,
J.B.\thinspace Kinson$^{   3}$,
A.\thinspace Kirk$^{   3}$,
A.A.\thinspace Kondashov$^{  5}$,
A.A.\thinspace Lednev$^{  5}$,
V.\thinspace Lenti$^{  4}$,
I.\thinspace Minashvili$^{   7}$,
J.P.\thinspace Peigneux$^{  1}$,
V.\thinspace Romanovsky$^{   7}$,
N.\thinspace Russakovich$^{   7}$,
A.\thinspace Semenov$^{   7}$,
P.M.\thinspace Shagin$^{  5}$,
H.\thinspace Shimizu$^{ 10}$,
A.V.\thinspace Singovsky$^{ 1,5}$,
A.\thinspace Sobol$^{   5}$,
M.\thinspace Stassinaki$^{   2}$,
J.P.\thinspace Stroot$^{  6}$,
K.\thinspace Takamatsu$^{ 9}$,
T.\thinspace Tsuru$^{   8}$,
O.\thinspace Villalobos Baillie$^{   3}$,
M.F.\thinspace Votruba$^{   3}$,
Y.\thinspace Yasu$^{   8}$.
}\end{center}

\begin{center}{\bf {{\bf Abstract}}}\end{center}

{
The reaction
$ pp \rightarrow p_{f} (\pi^{0}\pi^{0}\pi^{0}) p_{s}$
has been studied at 450 GeV/c.
The $\pi^0\pi^0\pi^0$ effective mass spectrum shows
clear $\eta(547)$ and $\pi_2(1670)$ signals.
Branching ratios for the $\eta(547)$ and $\pi_2(1670)$
are given as well as upper limits for the decays of the
$\omega(782)$, $a_1(1260)$ and $a_2(1320)$ into
$3\pi^0$.
}
\bigskip
\bigskip
\bigskip
\bigskip\begin{center}{{Submitted to Physics Letters}}
\end{center}
\bigskip
\bigskip
\begin{tabbing}
aba \=   \kill
$^1$ \> \small
LAPP-IN2P3, Annecy, France. \\
$^2$ \> \small
Athens University, Physics Department, Athens, Greece. \\
$^3$ \> \small
School of Physics and Astronomy, University of Birmingham, Birmingham, U.K. \\
$^4$ \> \small
CERN - European Organization for Nuclear Research, Geneva, Switzerland. \\
$^5$ \> \small
IHEP, Protvino, Russia. \\
$^6$ \> \small
IISN, Belgium. \\
$^7$ \> \small
JINR, Dubna, Russia. \\
$^8$ \> \small
High Energy Accelerator Research Organization (KEK), Tsukuba, Ibaraki 305-0801,
Japan. \\
$^{9}$ \> \small
Faculty of Engineering, Miyazaki University, Miyazaki 889-2192, Japan. \\
$^{10}$ \> \small
RCNP, Osaka University, Ibaraki, Osaka 567-0047, Japan. \\
$^{11}$ \> \small
Oslo University, Oslo, Norway. \\
$^{12}$ \> \small
Faculty of Science, Tohoku University, Aoba-ku, Sendai 980-8577, Japan. \\
\end{tabbing}
\end{titlepage}
\setcounter{page}{2}
\bigskip
In a previous analysis of the centrally produced \pipipic final state
clear signals of the $\eta(547)$, $\omega(782)$, $a_1(1260)$, $a_2(1320)$ and
$\pi_2(1670)$ were observed~\cite{3pipap}.
In particular, the $a_1(1260)$ and $a_2(1320)$ were
observed to decay dominantly to
$\rho \pi$. The $\pi_2(1670)$ was observed to decay to $\rho \pi$ and
$f_2(1270)\pi$. In order to gain more information on the
decays of these states it would be interesting to
study the \pipipi final state since only isospin zero isobars can
contribute to this final state.
\par
This paper presents new results from the WA102 experiment on
the centrally produced
\pipipi final state in the reaction
\begin{equation}
pp \rightarrow p_{f} (\pi^{0}\pi^{0}\pi^{0}) p_{s}
\label{eq:a}
\end{equation}
at 450 GeV/c.
The subscripts $f$ and $s$ indicate the
fastest and slowest particles in the laboratory, respectively.
\par
The data come from experiment WA102
which has been performed using the CERN Omega Spectrometer,
the layout of which is described in ref.~\cite{WADPT}.
\par
Reaction~(\ref{eq:a})
has been isolated from the sample of events having two outgoing
charged tracks plus six $\gamma$s 
reconstructed in the GAMS-4000
calorimeter,
by first imposing the following cuts on the components of the
missing momentum:
$|$missing~$P_{x}| <  14.0$ GeV/c,
$|$missing~$P_{y}| <  0.20$ GeV/c and
$|$missing~$P_{z}| <  0.16$ GeV/c,
where the x axis is along the beam
direction.
A correlation between
pulse-height and momentum
obtained from a system of
scintillation counters was used to ensure that the slow
particle was a proton.
\par
Fig.~\ref{fi:1} shows the two photon mass spectrum (5606 events)
for $6\gamma$-events  when
the mass of the other two $2\gamma$-pairs lies
within a band around the $\pi^0$ mass (100--170 MeV).
A clear $\pi^0$ signal is observed with a small background.
Events belonging to
reaction (\ref{eq:a}) have been selected using a
kinematical fit (7C fit, four-momentum
conservation being used and the masses of three $\pi^0$s being fixed).
The major background to the \pipipi final state comes from the
decay of the $\eta^\prime$ and $f_1(1285)$ to $\eta \pi^0 \pi^0$.
A kinematical fit has been used to remove these events.
The final sample consists of 3590 events and has less than 2 \%
contamination from the $\eta \pi^0 \pi^0 $ final state. The combinatorial
background is also reduced by the fact that only the
combination with the lowest $\chi^2$ is retained.
\par
Fig.~\ref{fi:2} shows the acceptance corrected
\pipipi
effective mass spectrum renormalised to the total number of observed events.
In addition to a clear $\eta(547)$
signal there is a broad enhancement
which is probably due to the $\pi_2(1670)$.
\par
The \pipipi mass spectrum shown in
fig.~\ref{fi:2} has been fitted with
a Gaussian~($\sigma$~=~16MeV) to describe
the $\eta(547)$, a Breit-Wigner
convoluted with a Gaussian~($\sigma$~=~32~MeV) to describe the $\pi_2(1670)$
and a
background of the form
$a(m-m_{th})^b$~exp$(-cm-dm^2-em^3)$, where
$m$ is the \pipipi mass,
$m_{th}$ is the threshold mass and
a,b,c,d,e are fit parameters.
The fit is found to describe the data
well and yields masses for the $\eta(547)$ and $\pi_2(1670)$ of:
\[
m(\eta(547)) \;\; = \;\; 545 \;\; \pm 0.6 \;\; \pm 0.5 \;\;{\rm MeV,}
\]
\[
m(\pi_2(1670)) \;\; = \;\; 1685 \;\; \pm 10  \;\; \pm 30 \;\;{\rm MeV}
\]
and
\[
\Gamma(\pi_2(1670)) \;\; = \;\; 265 \;\; \pm 30  \;\; \pm 40 \;\;{\rm MeV.}
\]
\par
A Dalitz plot analysis
of the \pipipi final state has been performed
using Zemach tensors and a standard isobar model~\cite{ABRAM}.
The analysis has assumed $\sigma$, $f_0(980)$, $f_2(1270)$ and $f_0(1370)$
intermediate states
and that
only relative angular momenta up to 2 contribute.
The $\sigma$ stands for the $\pi\pi$ S-wave amplitude squared,
and the parameterisation of
Zou and
Bugg~\cite{re:zbugg} has been used in this analysis.
\par
The geometrical acceptance of the apparatus has also been evaluated over
the Dalitz plot of the \pipipi system
in 40~MeV intervals between 0.8 and 2.0 GeV.
In order to perform a spin parity analysis the
log likelihood function, ${\cal L}_{j}=\sum_{i}\log P_{j}(i)$,
is defined by combining the probabilities of all events in 40 MeV
\pipipi mass bins from 0.80 to 2.0 GeV.
In order to include more than one wave in the fit
the incoherent sum of various
event fractions $a_{j}$ is calculated:
\begin{equation}
{\cal L}=\sum_{i}\log \left(\sum_{j}a_{j}P_{j}(i) +
(1-\sum_{j}a_{j})\right)
\end{equation}
where the term
$(1-\sum_{j}a_{j})$ represents the phase space background
which is a free parameter in each bin.
The negative log likelihood function ($-{\cal L} $) is then minimised using
MINUIT~\cite{re:MINUIT}.
Different combinations of waves and isobars have been tried and
insignificant contributions have been removed from the fit.
The fit generates the phase space background as that part of the data
not associated with a given wave
on a bin by bin basis and one requirement is that this
background is a smoothly varying function that shows no residual
resonance structure.
\par
Above 0.8 GeV only the addition of the
$J^{PC}=2^{-+}$~$f_2(1270) \pi^0$ S-wave produces a significant
change in the likelihood.
We can not exclude up to 3 \% contribution of
the $J^{PC}=2^{-+}$~$f_2(1270) \pi^0$ D-wave in the
$\pi_2(1670)$ region. Nor can we exclude that up to
10 \% of the $\pi_2(1670)$ comes from a $\sigma \pi$ final state.
The addition of any $1^{++}$ wave changes the log likelihood by
less than 1 unit and hence is classed as insignificant.
However, in order to estimate an upper limit on
the amount of $a_1(1260)$ decaying to \pipipi we include
both the $1^{++} \sigma \pi $ P-wave and the $1^{++} f_0(1370) \pi$ P-wave
in the final fit. The results of the final fit are shown in
fig.~\ref{fi:3}.
The
$J^{PC}=2^{-+}$~$f_2(1270) \pi^0$ S-wave well describes the peak in the
$\pi_2(1670)$ region. The $1^{++}$ wave is small and flat.
\par
Using the acceptance corrected number of events
from fits to the
\pipipi and \pipipic~\cite{3pipap} mass spectra
the branching ratio for the $\eta(547)$ to \pipipi and $\pi^+\pi^-\pi^0$
has been calculated
to be:
\[
\frac{\eta(547) \rightarrow \pi^0\pi^0\pi^0}
{\eta(547) \rightarrow \pi^+\pi^-\pi^0} = 1.35  \pm 0.06 \pm 0.09
\]
which is in agreement with the PDG value~\cite{PDG00} of 1.40~$\pm$~0.03.
The branching ratio for the $\pi_2(1670)$ to \pipipi and $\pi^+\pi^-\pi^0$
has been calculated in a similar manner to be:
\[
\frac{\pi_2(1670) \rightarrow \pi^0\pi^0\pi^0}{\pi_2(1670) \rightarrow
\pi^+\pi^-\pi^0} = 0.29  \pm 0.03 \pm 0.05
\]
\par
There is no evidence for a \pipipi decay mode of the $\omega(782)$, $a_1(1260)$
or $a_2(1320)$ therefore an upper limit has been calculated.
The masses and widths determined from the fit to the \pipipic channel
have been convoluted with the experimental resolution for the \pipipi final
state. The number of events, $N$, within 90 \% of the predicted resonance
profile has been determined and the upper limit has been calculated
using $2\sqrt{N}$.
For the $\omega(782)$
\[
\frac{\omega(782) \rightarrow \pi^0\pi^0\pi^0}
{\omega(782) \rightarrow \pi^+\pi^-\pi^0}
< 9 \times 10^{-4} \;\; (90 \% \;\;CL)
\]
to be compared with the PDG upper limit of $3 \times 10^{-4}$~\cite{PDG00}
which came from one experiment.
\par
For the $a_1(1260)$
\[
\frac{a_1(1260) \rightarrow \pi^0\pi^0\pi^0}
{a_1(1260) \rightarrow \pi^+\pi^-\pi^0}
< 8 \times 10^{-3} \;\; (90 \% \;\;CL)
\]
if on the other hand we used the total number of events in the
$1^{++}$ wave we would get:
\[
\frac{a_1(1260) \rightarrow \pi^0\pi^0\pi^0}
{a_1(1260) \rightarrow \pi^+\pi^-\pi^0}
< 9 \times 10^{-3} \;\; (90 \% \;\;CL)
\]
This imposes tight constraints on the decay of the
$a_1(1260)$ to isobars which have isospin 1. In particular,
it is in disagreement with
the claimed observation  of
$\sigma \pi$, $f_0(1370)\pi$ and $f_2(1270)\pi$
decay modes of the $a_1(1260)$ in ref.~\cite{ASNER}.
This claim was based on a Dalitz plot analysis of the
$a_1(1260)$ observed in $\tau$ decays.
Combining all the I~=~0 decays claimed in ref.~\cite{ASNER}
we have calculated the number of events we would expect to observe in the
\pipipi mass spectrum based on the number of $a_1(1260)$ observed in the
\pipipic final state of experiment WA102~\cite{3pipap}.
Superimposed on the \pipipi mass
spectrum in fig.~\ref{fi:4} is the $a_1(1260)$ signal we would expect to
observe based on this number of events.
As can be seen, irrespective of any results from the spin analysis,
the number of events in the \pipipi spectrum is much smaller
than the predicted signal. The fact that the $\eta$ branching ratio
we have measured is in agreement with the PDG value indicates that our
normalisation between the \pipipi and \pipipic channels is correct. Therefore
this discrepancy in the possible I=0 decay modes of the $a_1(1260)$ could
either
be due to an overestimate of the number of $a_1(1260)$ events in the
\pipipic final state of the WA102 experiment or due to an
error in the spin analysis performed by CLEO in ref.~\cite{ASNER}.
\par
Finally, for the $a_2(1320)$ we obtain
\[
\frac{a_2(1320) \rightarrow \pi^0\pi^0\pi^0}
{a_2(1320) \rightarrow \pi^+\pi^-\pi^0}
< 9 \times 10^{-3} \;\; (90 \% \;\;CL)
\]
\par
In summary, a study of the centrally produced $\pi^0\pi^0\pi^0$ system
shows prominent signals of the $\eta(547)$ and $\pi_2(1670)$.
Branching ratios for the $\eta(547)$ and $\pi_2(1670)$ are given.
Upper limits are calculated for the $\omega(782)$, $a_1(1260)$
and $a_2(1320)$ which can be used to constrain the possible
decays of these states to isobars with isospin zero.
\bigskip
\begin{center}
{\bf Acknowledgements}
\end{center}
\par
This work is supported, in part, by grants from
the British Particle Physics and Astronomy Research Council,
the British Royal Society,
the Ministry of Education, Science, Sports and Culture of Japan
(grants no. 1004100 and 07044098), the French Programme International
de Cooperation Scientifique (grant no. 576)
and
the Russian Foundation for Basic Research
(grants 96-15-96633 and 98-02-22032).

\newpage
{ \large \bf Figures \rm}
\begin{figure}[h]
\caption{
The M($\gamma \gamma$) when the other two $\gamma \gamma$ pairs lie
in the $\pi^0$ mass region.
}
\label{fi:1}
\end{figure}
\begin{figure}[h]
\caption{The
\pipipi
effective mass spectrum,
with fit described in the text.}
\label{fi:2}
\end{figure}
\begin{figure}[h]
\caption{Results of the spin parity analysis.}
\label{fi:3}
\end{figure}
\begin{figure}[h]
\caption{The
\pipipi
effective mass spectrum,
with superimposed the number of events expected from
the $a_1(1260)$, see text.}
\label{fi:4}
\end{figure}
\newpage
\begin{center}
\epsfig{figure=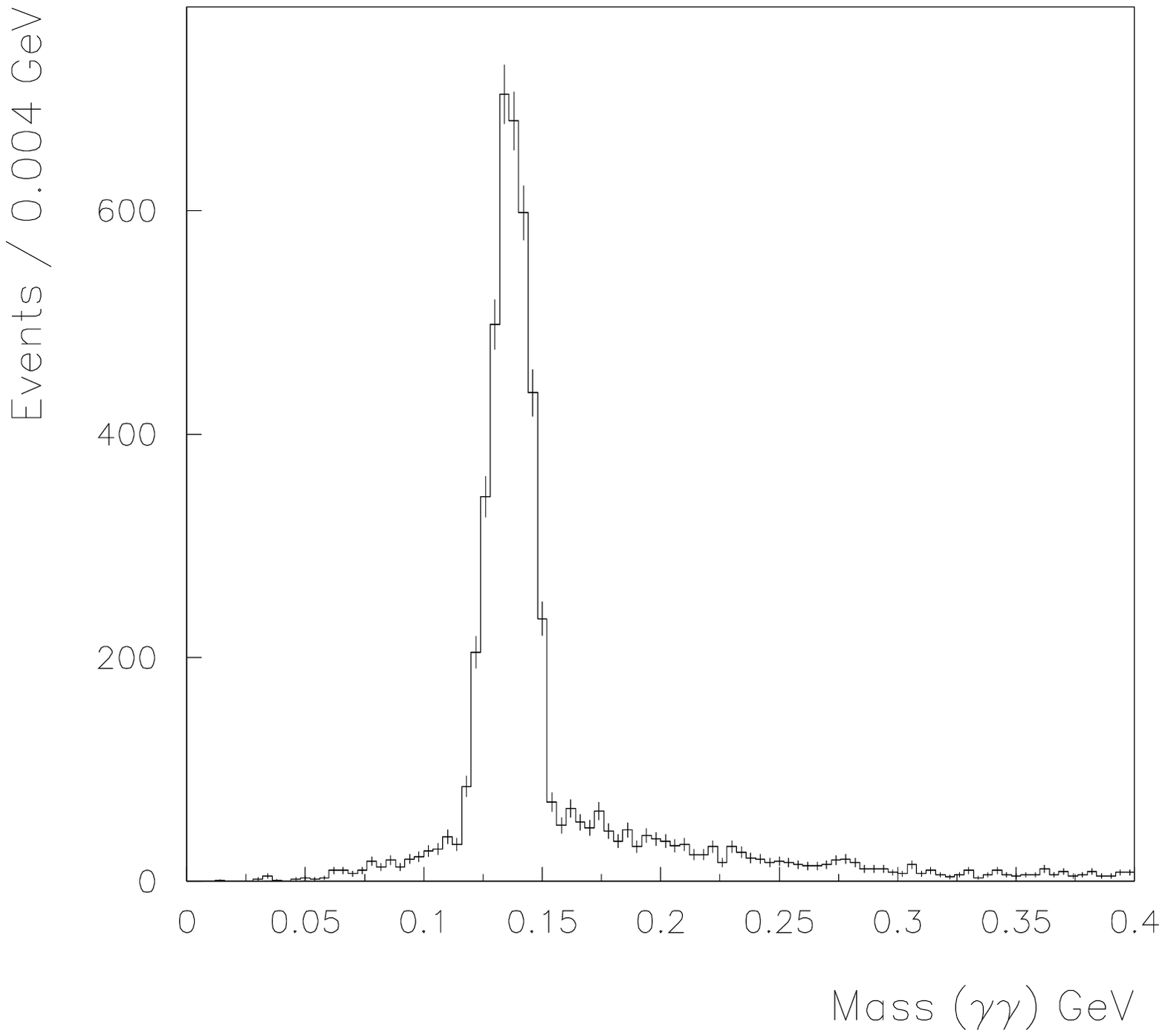,height=22cm,width=17cm}
\end{center}
\begin{center} {Figure 1} \end{center}
\newpage
\begin{center}
\epsfig{figure=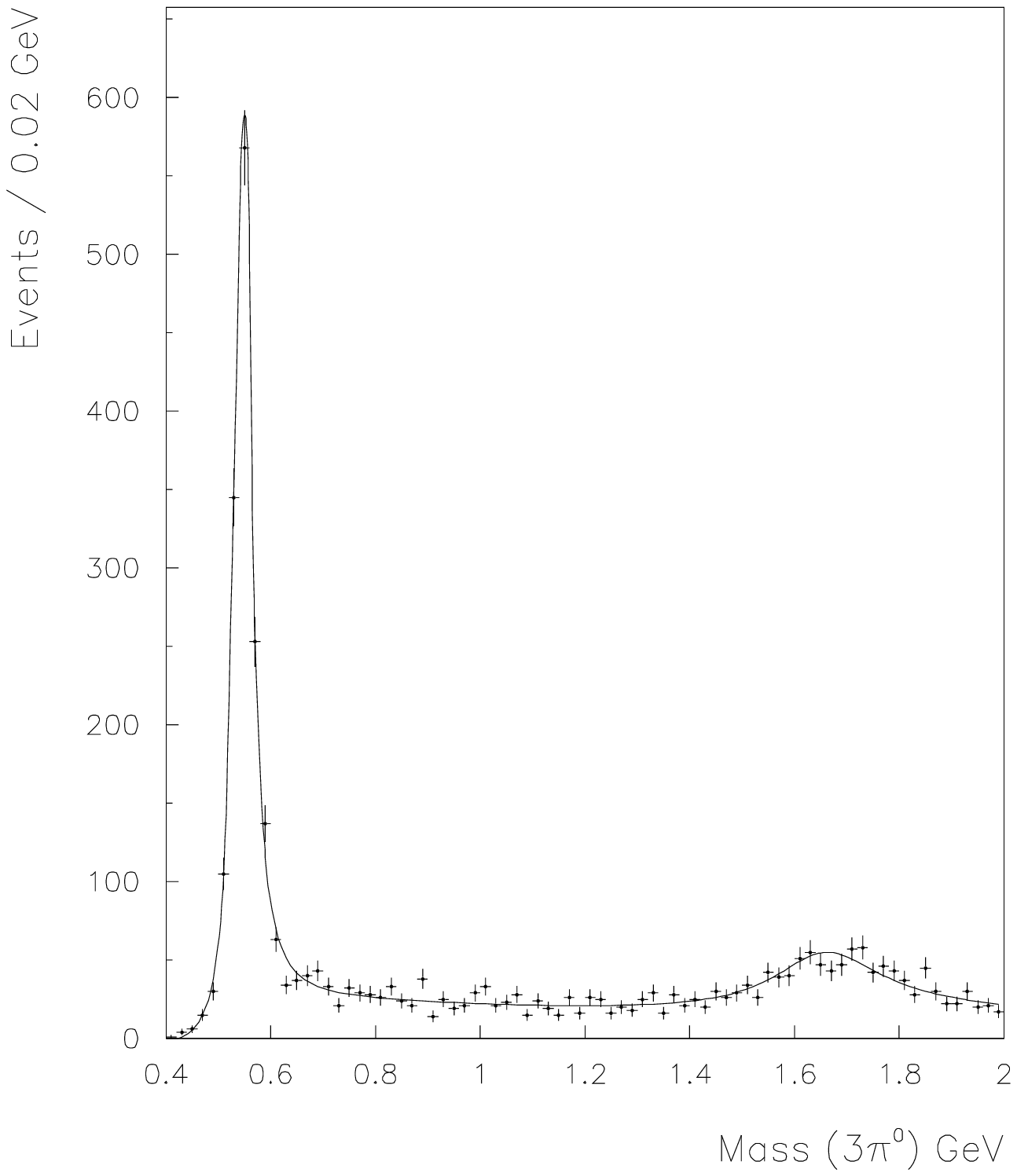,height=22cm,width=17cm}
\end{center}
\begin{center} {Figure 2} \end{center}
\newpage
\begin{center}
\epsfig{figure=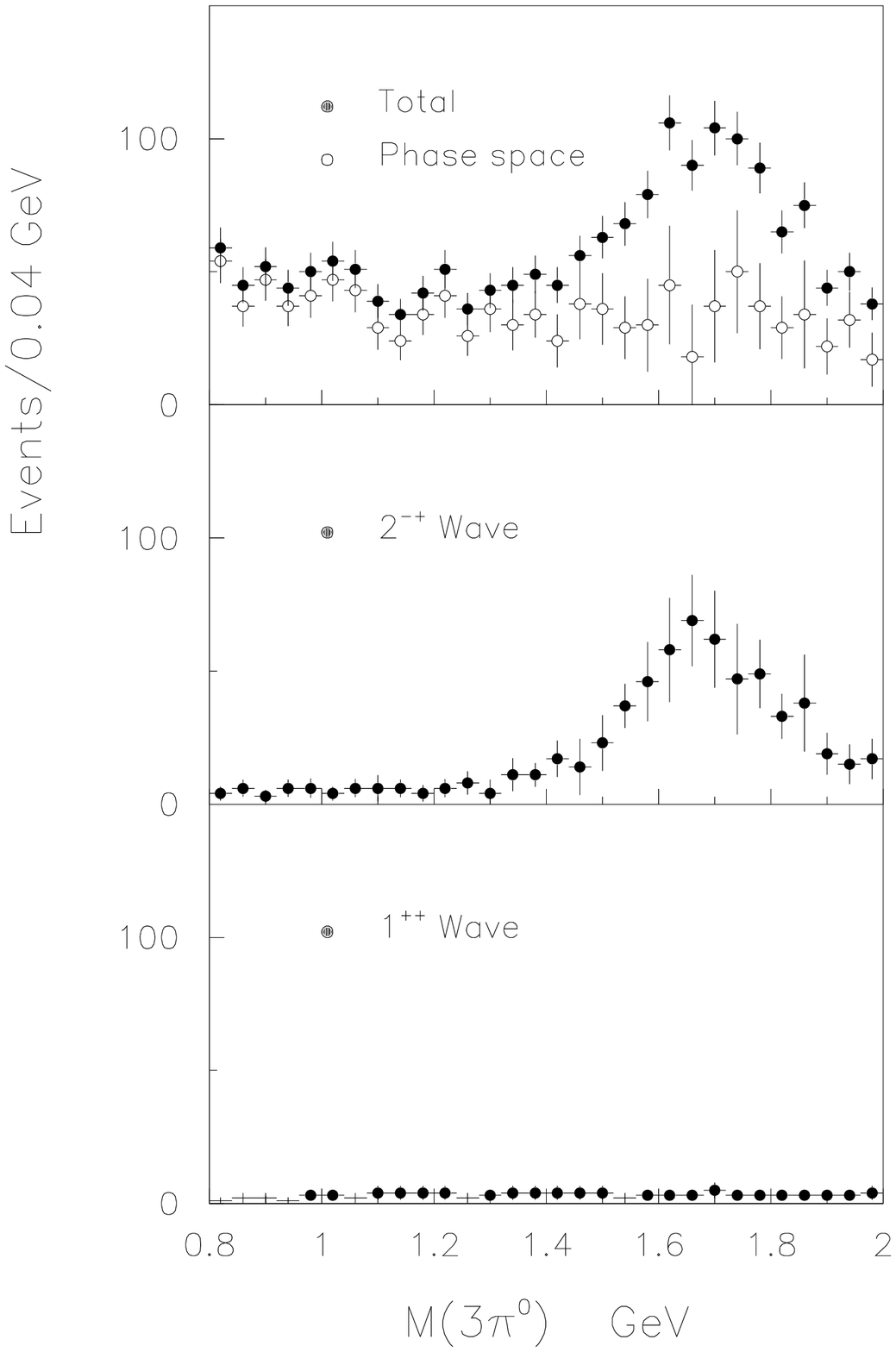,height=20cm,width=17cm,
bbllx=0pt,bblly=0pt,bburx=680pt,bbury=650pt}
\end{center}
\begin{center} {Figure 3} \end{center}
\newpage
\begin{center}
\epsfig{figure=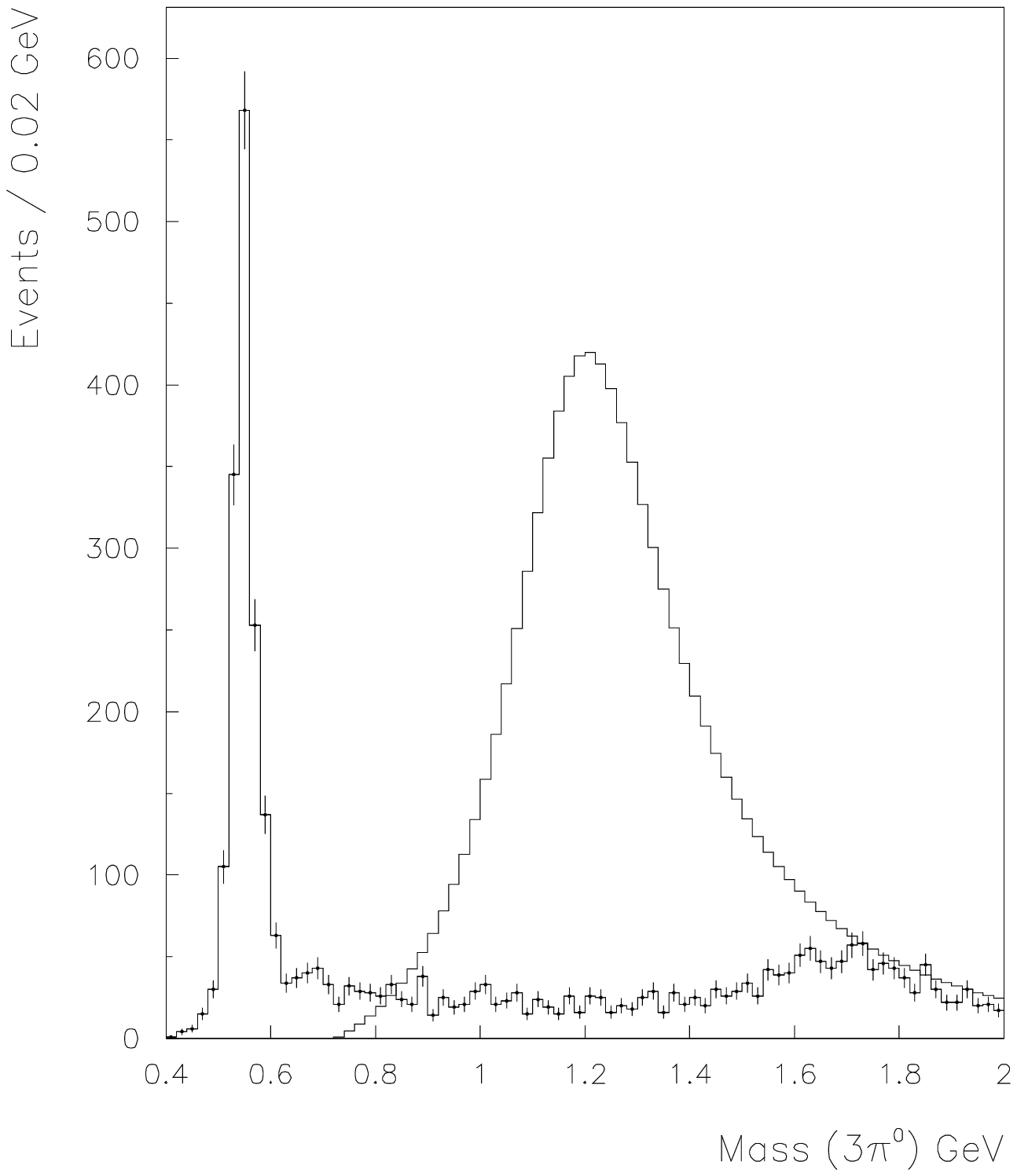,height=22cm,width=17cm}
\end{center}
\begin{center} {Figure 4} \end{center}
\end{document}